# Metal to insulator transition in Conducting Polyaniline/Graphene Oxide composites

Eleni Neti [1], Elias Sakellis [1,2], Anthony N. Papathanassiou [1,*],

Evangelos Vitoratos [3] and Sotirios Sakkopoulos [3]

[1] National and Kapodistrian University of Athens, Physics Department. Panepistimiopolis, GR 15784 Zografos, Athens, Greece

[2] National Center of Natural Sciences Demokritos, Institute of Nanomaterials and Nanotechology, Aghia Paraskevi, Athens, Greece

[3] University of Patras, Physics Department, 265 00 Patras, Greece

**Abstract**

Broadband Dielectric Spectroscopy (BDS) measurements of Polyaniline/Graphene oxide composites were conducted for an as-prepared and a thermally annealed specimen, respectively, from 15K to room temperature. Electrical conductivity values of the annealed composite display a very modest rise denoting the important contributions of the GO component to achieving electrical stability of the polymer. Patterns of the dc conductivity as a function of temperature also reveal a metal to insulator transition around 75K. The transition is dominated by two key factors; temperature and annealing process. Metal-like and insulating features are subsequently detected, as well, and accordingly described to provide a qualitative inspection of the charge transfer mechanisms involved.

## 1. Introduction

The rapidly emerging field of polymer science has introduced many innovative nanocomposites of significant technological and scientific importance. Indisputably, in modern technology, novel materials that combine superior electric properties, stability and low cost production are highly desired. Polyaniline (PANI), as an intrinsically conducting polymer [1], is a very promising candidate for developing functional graphene/polymer composites. PANI possesses exceptional physicochemical properties, such as flexibility, solution processability and tunable conductivity by undergoing reversible doping processes (protonation mechanism of imine nitrogen atoms) [2]. Graphene oxide (GO) is considered an attractive graphene-based material produced in large scale and at a lowest cost [10]. Similarly to graphene, GO presents a $sp^2$ hybridized planar structure except for diverse chemical oxygen-based groups present on the surface. These functional groups render GO itself electrically insulating and thermally unstable but also provide a substrate for a variety of chemical modifications and chemical bonds [3,4].

It is well known that protonation of PANI results in its reduction to a disordered system with a granural metal type structure involving regions of metallic conductivity randomly interspersed within the amorphous phase [5]. The metallic phases originate from the highly-packed ordered chains that form metallic bundles or "grains" in which electron states are three-dimensionally extended [6]. The macroscopic conductivity is then dictated either by quasi-1D hopping [5] or fluctuation-induced tunnelling [24] of charge carriers penetrating the potential barrier that separates neighboring metallic domains. With regard to the material under investigation in the present work, dc conductivity dependence upon temperature reveals a metal to insulator transition [7]. In general, disorder in organic "quasi-metals" arises from a combination of microscopic disorder as well as structural inhomogeneity at mesoscopic scales [8, 9]. The term "metallic" is broadly used to describe polymers that share a common characteristic with conventional metals, whereas finite resistivity $\rho(T)$ is observed as an effect of the random collisions of the electrons drifting ballistically through the metal, even when temperature approaches absolute zero [8].

To the best of our knowledge, metallic behavior in PANI/GO, also expanding in a temperature range up to 75 K, has not yet been reposted. Considering that PANI/GO has been extensively investigated as a material for organic electronics technology due to its high capacitance and electrochemical performance by easy reversal doping or dedoping processes [10-13], here we focus on a conducting PANI composite with enhanced electrical properties targeted to semiconductor applications. The aim of this study is to enlighten the physical phenomena that arise in the microscopic scale and contribute to the aforementioned crossover in nature of the dc conductivity. The stability of the dc conductivity and its transition temperature are comparatively explored and interpreted.

## 2. Experimental details

In the present work polyaniline (PANI) and two different nanocomposites of polyaniline with graphene oxide (GO) were prepared by adding 1.5 and 3% w/w of the latter in the polymer. PANI was prepared from the monomer by in – situ polymerization with ammonium persulfate $(NH_4)_2S_2O_8$ as oxidant in the presence of 3D $H_2O$/HCl 1M (aniline/APS 1/1). In the case of nanocomposites the proper amount of grapheme oxide was added in the solution and 1h of sonication ensured homogeneous distribution of the oxide. Teflon or polycarbonate filters were proved suitable for the isolation of the final product in the form of dark powder, which was dried and pressed in circular pellets about 1.3 mm in diameter and 1 mm thick [14-16].

The specimens were placed inside a capacitor type sample holder of a high-vacuum closed-circuit liquid helium cryostat (ROK, Leybold-Heraeus) operating from 15 K to room temperature. Temperature was stabilized with an accuracy of 0.01 K using a LTC 60 temperature controller. Dielectric measurements were performed in the frequency range of 1 mHz to 1 MHz with a Solartron SI 1260 Gain-Phase Frequency Response Analyzer and a Broadband Dielectric Converter (BDC, Novocontrol). WinDeta (Novocontrol) software was used for monitoring the data acquisition [17].

## 3. Results and discussion

### *Distinction among different conductivity modes*

Within the context of a transition from a metallic to an insulating behavior, in the vast majority of conducting polymers, dc-conductivity vs. temperature plots can be divided into three distinct regimes [18]; i) in the insulating regime, where conductivity exhibits a positive temperature dependence and extrapolates to zero at low *T*, Mott-like behaving [19] ii) in the critical regime, where the resistivity is *approximately* not activated [20] iii) in the metallic regime where the conductivity exhibits a negative temperature coefficient and remains non-zero at low T.

In principle, complex conductivity $\sigma^*(f)=i\omega\varepsilon_o\varepsilon^*(f)$, where $\varepsilon_o$ denotes the permittivity of vacuum and $\varepsilon^*(f)$ is the complex permittivity, consists of a real and an imaginary part, whereas the real part $\sigma'(f)$ includes a constant dc component $\sigma_{dc}$ [21]. The BDS spectra recorded (Fig.1) indicate that the overall electrical conductivity of the blends dominates to any dielectric relaxation peak that is probably stimulated by interfacial polarization processes due to the inhomogeneous structure. As can be seen in the aforementioned typical diagram, σ´(f) isotherms from 16K to the room temperature limit for PANI(APS)/GO, are frequency-independent and remains so for the entire frequency range employed in this work. The temperature dependence of the dc conductivity $\sigma_{dc}$, depicted in Fig. 2, is characterized by a variation in its monotony at low temperatures. The latter can be effectually described in terms of a Metal to Insulator transition (MIT). The slope for the *fresh* sample exhibits a sign reversal around 75 K, suggesting that a MIT occurs. Concerning the *annealed* sample, the slope retains its sign, but a small change in the (absolute) value, around the same temperature reported above for the fresh sample, evidences for a crossover to the insulating regime; however, it seems that the transition spans over a broad temperature range below 75 K.

Quantum phase transitions in conductivity-temperature transitions are accurately determined by plotting the Stickel plots [22]. A Stickel plot consists of the logarithm of $\sigma_{dc}$ versus reciprocal temperature (Fig. 3). Straight lines can fit the data points below and above 75 K, respectively, the slope of which provide *apparent* activation energy values $E^{act} \equiv -2.3\, dlog\sigma_{dc}/d(1/kT)$, where k is the Boltzmann constant. While positive value of $E^{act}$ is related to the effective height of a potential barrier separating randomly distributed conducting regions, negative value signifies the "barrier-less" motion, i.e., quasi-free electron transport. Thus, the metal-insulator transition is manifested by the sign reversal of $E^{act}$.

*The effect of GO on the transition temperature*

The conductivity-temperature plot of *neat* PANI(APS) (inset diagram Fig.2) reveals the absence of any transition above 15 K and, moreover, displays a semiconducting behavior in the entire temperature range studied by increasing smoothly with temperature. Subsequently, the transition observed around 75 K in PANI(APS)/GO results from the accommodation of GO in favor of a conduction network consisting of conducting "pathways" or regions. Because of its graphene-like nature, GO possesses a wealth of extended electronic states readily introduced into the transport process. Due to their abundance, electron-electron interactions are facilitated and retain this interaction up to a MIT temperature higher than that of *neat* PANI (APS). Within the low-temperature metallic regime the *fresh* PANI(APS)/GO sample has delocalized electronic states at the Fermi level boosting the overall electrical conductivity, and conduction is permitted without thermal activation, hence by traversing the grainy network as if in a continuous "barrier-less" landscape. The transition towards the metallic state upon cooling has been traced previously for many different conducting polymers, as well. For instance, for $PF_6$-doped poly(3,4-ethylenedioxythiophene), its conductivity decreases monotonically down to 10 K; further below, its temperature derivative changes its sign [23]. Such behavior is common in amorphous metals, and the transition is interpreted as the low-temperature contribution of electron-electron interactions [18], i.e. backscattered electronic wavefunctions interfere augmenting resistance at low temperatures.

Above the MIT temperature, both fresh and thermally treated PANI(APS)/GO samples exhibit a common $T^{-1/2}$ dependence (Fig.4). The granular-metal-like fluctuation induced tunnelling model (FIT) [24], predicts a dependence:

$$\sigma_{dc}(T) \propto exp[-(T_o/T)^{1/2}] \qquad (1)$$

where $T_0$ being a fitting parameter linking to an apparent activation energy $E_{a,0} = kT_0$. The aforementioned dependence picture is dominant at higher temperatures (T >75 K) and evidences for a phonon-assisted-tunnelling mechanism of carriers transferring among the conductive grains [25]. We note that a temperature dependence similar to eq.(1) is implied alternatively for conducting PANI within the charging energy limited tunnelling model (CELT) [26]. Succinctly, the insulating phase is well manifested above the transition point, as the quasi-free electronic transport is obstructed by the increased vibrational motion of atoms and molecules and the electron-electron contributions become negligible.

The values of the characteristic parameter of eq. (1) $T_o$ are obtained from the slope of the plots depicted in Fig.4. Its increase from $T_o=3.1(1)$ K to $T_o=11.5(2)$ K induced by the annealing process, is commonly associated with the reduction of the conductive protonated phase and the shrinking of the metallic grains in conducting PANI systems followed by thermal aging [27-31]. This is also the case for conducting PANI(APS)/GO; The parameter $T_o$ is directly related to the ratio $s/d$ as:

$$\frac{s}{d} = \frac{kT_o}{16U}\left[1 + \sqrt{1 + \frac{16U}{kT_o}}\right] \qquad (2)$$

where $s$ is the average distance between neighboring conductive grains, $d$ is the average grain size and $U$ denotes the Coulomb repulsive energy of two electrons at a distance equal to the size of a PANI monomer [28]. Given that U is roughly 2 eV and to retain its value constant as a function of aging, the ratio $s/d$ is found to increase almost as 93%; according to our calculations based on eq. (2), the ratio increases from $s/d \approx 0.0029$ to $s/d \approx 0.0056$. The remarkable increase of the $s/d$ ratio reflects the striking reduction of the grain size ($d$) while the separating barrier width ($s$) follows the opposite trend. The shrink of conducting grains after isothermal anneal at 70º C for 2 h is in accordance with the increase of the $E^{act}$ values by approximately 4 times, as shown in Fig. 3, indicating that the macroscopic charge conduction faces higher effective potential barriers alongside an annealed polymer matrix. From this point of view, the resultant broadening of the insulating regions in expense of the metallic ones, accounts for substantial disorder which disturbs the extended conduction network and causes the electronic states to become localized. Obviously, annealing plays a revealing role to achieving intense semiconducting state, and therefore extended charge motion is suppressed and the MIT point shifts towards to a lower temperature , i.e., below 15K. However, one can observe weaker temperature dependence of the dc conductivity in spanning from 75 K to 15 K (Fig.2): the conductivity is still dominated by phonon-assisted hopping between localized states, which competes with the extended electronic transport. While the thermal degradation of the electrical conductivity in various conducting PANI systems is reported in the literature [27-30], our findings in PANI(APS)/GO indicate a slight change of the net values of the dc conductivity. This modest increase can be explained as the effect of the drastic conductivity degradation observed in PANI counterbalanced by the systematic charge carrier supply from GO conducting islands. The latter, also constitutes a favorable indication that the introduction of GO plays a leading role for monitoring the concentration of electric charge carriers in the composite and optimizing the stability of the electrical properties of PANI as well.

## 4. Conclusions

The dc conductivity of doped-Polyaniline/Graphene oxide composites exhibits various temperature dependencies from 15 K to room temperature. The results are consistent with the picture of a heterogeneously disordered matrix resembling that of a granural metal. The grainy network is further enriched in extended electronic states supplied by the embedded GO. A transition from the metallic to the insulating phase is observed around 75 K and confirmed by plotting the Stickel diagrams, from which the effective height of potential barrier is determined. Thermal annealing alters the low temperature metallic region to a semi-insulating one by reversing the sign of the temperature derivative of the dc conductivity, probably as the result of the competition between the extended electronic transport and the localization effects due to the modification of the conduction network. Despite the aforementioned dynamics in electron transport after the thermal treatment, its net values are moderately increased, probably as the reduction of charge carriers due to deprotonization of PANI is balanced by those gained by the GO dispersion. GO is likely to maintain the overall conductivity on thermal annealing, while the carriers' dynamics departs from that of quasi-free electron transport.

**Acknowledgements**: We want to thank Dr. Spyros N.Yannopoulos (Research Director, Institute of Chemical Engineering Sciences, Foundation for Research and Technology Hellas (FORTH/ICE-HT), 26504Patras, Greece) and MSc Vasilios Petoumenos, for providing the specimens and fruitful discussions.

# References

[1] Molapo, K., P. Ndangili, R. Ajayi, G. Mbambisa, S. Mailu, N. Njomo, M. Masikini, P. Baker,and E. Iwuoha *Electronics of Conjugated Polymers (I): Polyaniline* **International Journal of Electrochemical Science 7**, pp. 11859- 11875 (2012)

[2] Huang, W.-S., Humphrey, B. D., & MacDiarmid, A. G. *Polyaniline, a novel conducting polymer. Morphology and chemistry of its oxidation and reduction in aqueous electrolytes.* **J. Chem. Soc., Faraday Trans. 1 82**, p.2385 (1986)

[3] Huang, X., Liu, F., Jiang, P., & Tanaka, T. *Is graphene oxide an insulating material?* **IEEE International Conference on Solid Dielectrics (ICSD)** (2013)

[4] Dreyer, D. R., Park, S., Bielawski, C. W., & Ruoff, R. S. *The chemistry of graphene oxide,* **Chem. Soc. Rev. 39,** pp. 228-240 (2010)

[5] Wang, Z. H., Scherr, E. M., MacDiarmid, A. G., & Epstein, A. J. *Transport and EPR studies of polyaniline: A quasi-one-dimensional conductor with three-dimensional ''metallic'' states* **Phys. Rev. B. 45**, pp.4190-4202 (1992)

[6] Wang, Z. H., Li, C., Scherr, E. M., MacDiarmid, A. G., & Epstein, A. J. *Three dimensionality of ''metallic'' states in conducting polymers: Polyaniline* **Phys. Rev. Lett. 66**, pp.1745-1748 (1991)

[7] N.E. Mott, *Metal-Insulator transitions* ( **Francis & Taylor,** London, 1990), pp.145-155

[8] Lee, K., Cho, S., Heum Park, S., Heeger, A. J., Lee, C.-W., & Lee, S.-H. *Metallic transport in polyaniline* **Nature 441 ,** pp. 65-68. (2006)

[9] Kaiser, A. B. *Systematic Conductivity Behavior in Conducting Polymers: Effects of Heterogeneous Disorder* **Adv. Mater. 13**, pp. 927-941 (2001)

[10] Vargas, L. R., Poli, A. K. D. S., Dutra, R. de C. L., De Souza, C. B., Baldan, M. R., & Gonçalves, E. S. *Formation of Composite Polyaniline and Graphene Oxide by Physical Mixture Method* **J.Aerosp.Technol.Manag. 9**, pp. 29-38 (2017)

[11] Wang, H., Hao, Q., Yang, X., Lu, L., & Wang, X. *Effect of Graphene Oxide on the Properties of Its Composite with Polyaniline* **ACS Appl. Mater. Interfaces 2**, pp. 821-828 (2010)

[12] Mombrú, D., Romero, M., Faccio, R., & Mombrú, Álvaro W. *Effect of graphene-oxide on the microstructure and charge carrier transport of polyaniline nanocomposites under low applied electric fields* **Journal of Applied Physics 121**, pp. 045109 (2017)

[13] S. Chauhan *Graphene Oxide/Polyaniline Composites as Electrode Material for Supercapacitors* **Journal of Chemical and Pharmaceutical Research 9**, pp. 285-291(2017)

[14] S. J. Lin, H. J. Sun, T. J. Peng, and L. H. Jiang *Synthesis of high-performance polyaniline/graphene oxide nanocomposites* **High Perform. Polym. 26**, pp. 790–797 (2014)

[15]A. Riaz, A. Usman, M. Faheem, Z. Hussain, A. N. Khan, and S. Soomro *Effect of polymerization of aniline on thermal stability, electrical conductivity and band gap of graphene oxide/polyaniline nanocomposites* **Int. J. Electrochem. Sci.12**, pp. 1785–1796 (2017)

[16] S. Konwer *Graphene oxide-polyaniline nanocomposites for high performance supercapacitor and their optical, electrical and electrochemical properties* **J. Mater. Sci. Mater. Electron.27**, pp. 4139–4146 (2016)

[17] Papathanassiou, A. N., Plonska-Brzezinska, M. E., Mykhailiv, O., Echegoyen, L., & Sakellis, I. *Combined high permittivity and high electrical conductivity of carbon nano-onion/polyaniline composites* **Synthetic Metals 209**, pp. 583-587 (2015)

[18] Le, T.-H., Kim, Y., & Yoon, H. *Electrical and Electrochemical Properties of Conducting Polymers* **Polymers 9**, p. 150 (2017)

[19] N.E. Mott and E.A. Davis, *Electronic processes in non-crystalline materials* (**Oxford University Press**, Oxford, 2012), pp.32-37

[20] Larkin, A., & Khmel'nitskii, D. *Activation conductivity in disordered systems with large lacalization length* **Sov.Phys.JETP 56** , p. 647 (1982)

[21] F. Kremer and A. Schönhals, *Broadband Dielectric Spectroscopy* (**Springer**, Berlin, 1991), p.81

[22] Stickel, F., Fischer, E. W., & Richert, R. *Dynamics of glass‐forming liquids. II. Detailed comparison of dielectric relaxation, dc‐conductivity, and viscosity data* **The Journal of Chemical Physics 104**, pp. 2043-2055 (1996)

[23] Aleshin, A., Kiebooms, R., Menon, R., Wudl, F., & Heeger, A. J. *Metallic conductivity at low temperatures in poly(3,4-ethylenedioxythiophene) doped with $PF_6$* **Phys. Rev. B 56,** pp. 3659-3663 (1997)

[24] Sheng, P. *Fluctuation-induced tunneling conduction in disordered materials* **Phys. Rev. B 21** , pp. 2180-2195 (1980)

[25] Sheng, P., Abeles, B., & Arie, Y. *Hopping Conductivity in Granular Metals.* **Phys. Rev. Lett. 31**, pp. 44-47 (1973)

[26] Zuppiroli, L., Bussac, M. N., Paschen, S., Chauvet, O., & Forro, L. *Hopping in disordered conducting polymers.* **Phys. Rev. B 50**, pp. 5196-5203 (1994)

[27] Dalas, E., Sakkopoulos, S., & Vitoratos, E. *Thermal degradation of the electrical conductivity in polyaniline and polypyrrole composites*   **Synthetic Metals 114**, pp. 365-368 (2000)

[28] Sakkopoulos, S., Vitoratos, E., Dalas, E., Kyriakopoulos, N., Malkaj, P., & Argyreas, T. *Shrinking rate of conducting grains in HCl-protonated polyaniline, polypyrrole, and polypyrrole/polyaniline blends with their thermal aging* **J. Appl. Polym. Sci. 97**, pp. 117-122 (2005)

[29] Sakkopoulos, S., Vitoratos, E., & Dalas, E *Conductivity degradation due to thermal aging in conducting polyaniline and polypyrrole*. **Synthetic Metals 92**, pp. 63-67 (1998)

[30] Jousseaume, V., Morsli, M., & Bonnet, A.   *Aging of electrical conductivity in conducting polymer films based on polyaniline.* **Journal of Applied Physics 88** , pp. 960-966 (2000)

[31] Rannou, P., Nechtschein, M., Travers, J., Berner, D., Woher, A., & Djurado, D. *Ageing of PANI: chemical, structural and transport consequences*. **Synthetic Metals 101** , pp. 734-737 (1999)

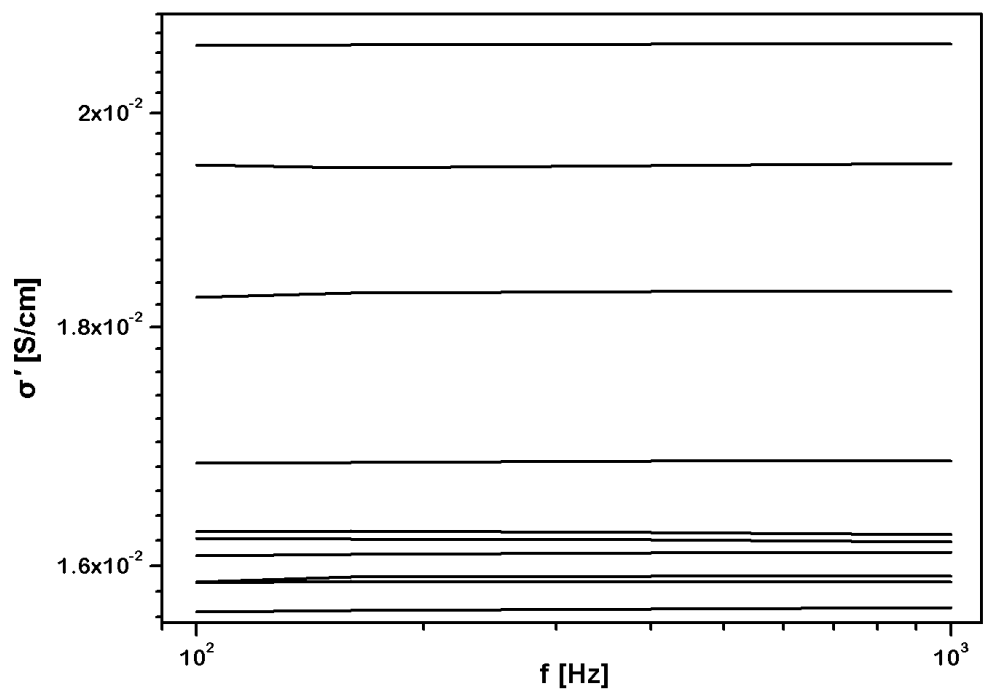

**Figure 1:** _A sample of $\sigma'(f)$ isotherms in a double logarithmic representation recorded for annealed PANI/GO composite. Temperature ranges from 290K to 16K, from top to bottom, respectively.

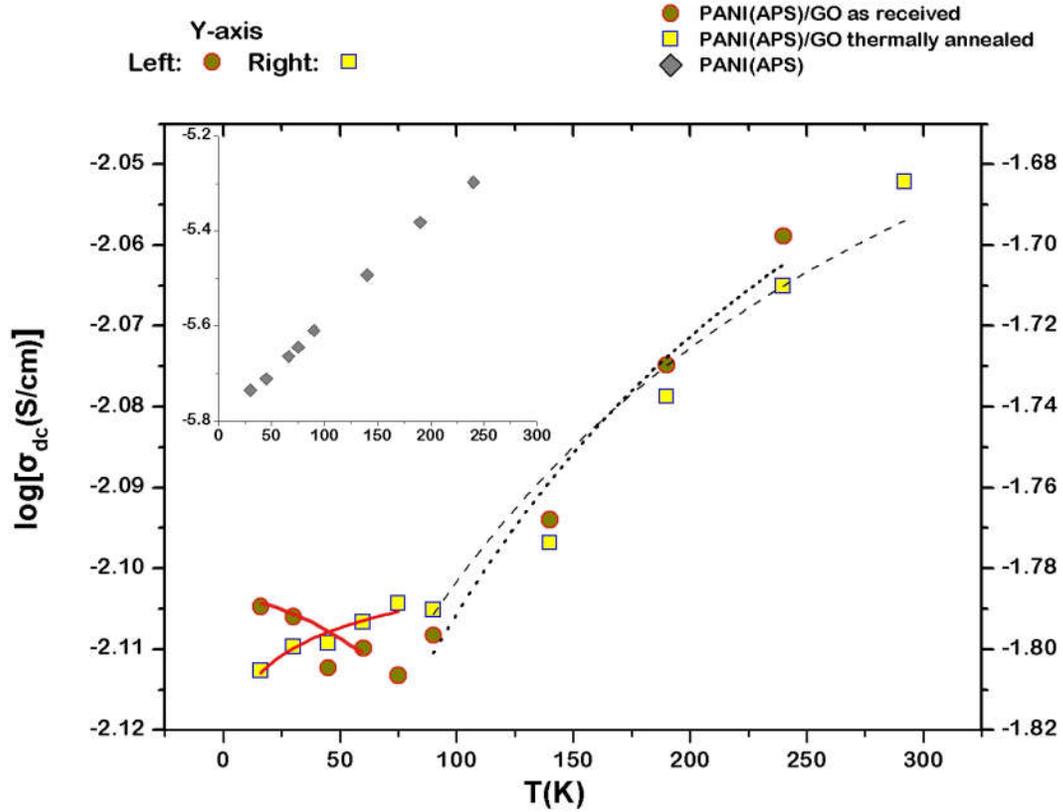

**Figure 2:** Temperature dependence of the dc conductivity for as-received and annealed PANI(APS)/1,5% w/w GO, respectively. A change in monotony around 75 K is observed. Dashed and dotted lines are drawn for a better visualization of the insulator area, Mott-like behaving. Straight lines account as empirical fitting function $\sigma_{dc}(T)=AT^b$; The fitting parameters are: b= 0.002(1), A= -2.091(8) for the *fresh* and b= -0.0053(9), A= -1.832(6) for the *annealed* specimen, respectively. In the inset diagram, dc conductivity-temperature plot recorded for neat PANI(APS) is depicted, whereas no monotony alteration is observed.

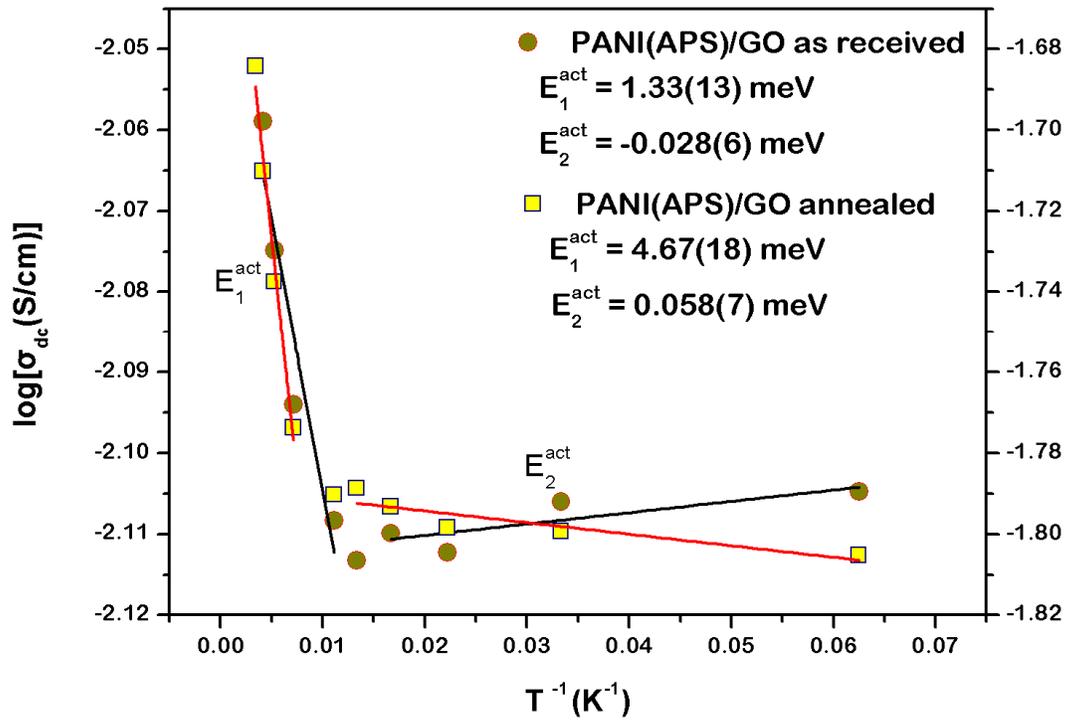

**Figure 3:** Stickel plot of the logarithm of dc conductivity indicates two distinct regions separated by a critical point determined to 75K. Evaluated $E^{act}$ energies obtained by linear fitting are included for each sample.

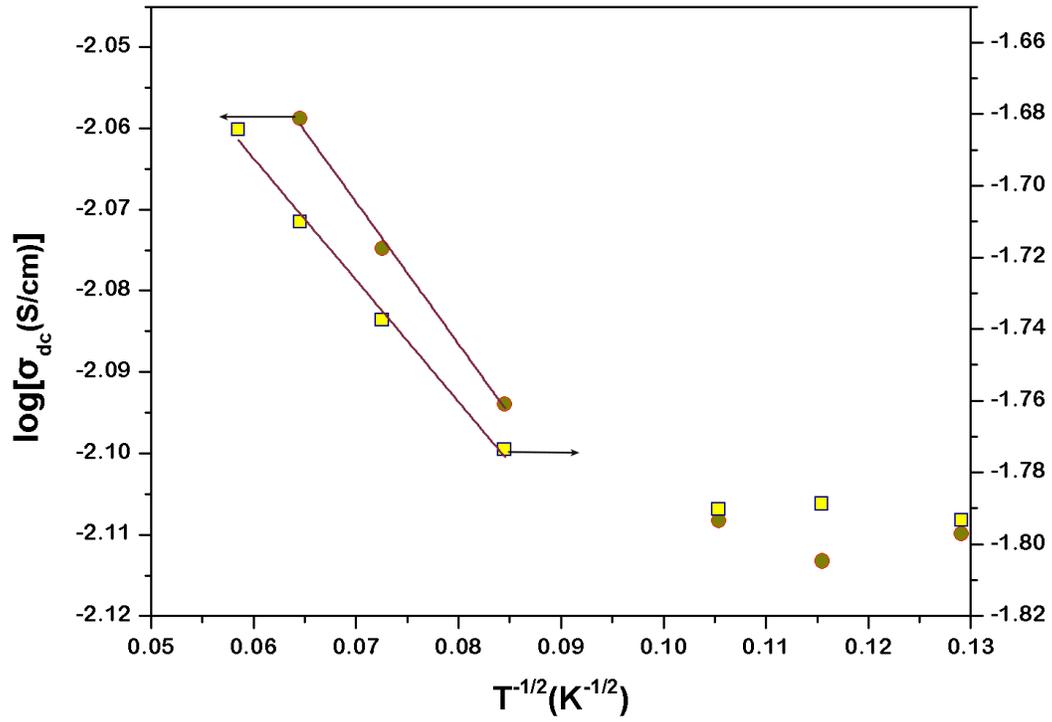

**Figure 4:** The dc conductivity as a function of $T^{-1/2}$ for PANI(APS) /1,5 (w/w) GO samples. For the insulating high-temperature regime (T>75K) data points lie on a straight line according to eq. (1). The fitting parameter *To* accounts as a measure of the impact of annealing in the heterogeneous system.